\documentclass[aps,prapplied,reprint,longbibliography,superscriptaddress]{revtex4-2}
\usepackage[T1]{fontenc}
\usepackage{textcomp}
\usepackage[utf8]{inputenc}
\usepackage{xcolor}
\usepackage{booktabs}
\usepackage{enumitem}
\usepackage{amsmath}
\usepackage{amssymb}
\usepackage{graphicx}
\PassOptionsToPackage{normalem}{ulem}
\usepackage{ulem}
\usepackage{microtype}
\usepackage[bookmarks=false,
    breaklinks=true,
    pdfborder={0 0 1},
    backref=false,
    colorlinks=true
 ]{hyperref}
 \hypersetup{
    citecolor=purple,
    linkcolor=purple,
    urlcolor=black
}
\usepackage[size=footnotesize]{todonotes}

\makeatletter

\providecommand{\tabularnewline}{\\}

\usepackage{orcidlink}
\usepackage{newtxtext,newtxmath}
\usepackage[english]{babel}

\def\fnum@figure{\textbf{Fig.~\thefigure}}
\def\fnum@table{\textbf{Tab.~\thetable}}

\makeatother

\begin{document}
\title{Optimization of Si/SiGe Heterostructures for Large and Robust Valley Splitting in Silicon Qubits}

\author{Abel Thayil \orcidlink{0000-0002-3438-5901}}
\email{thayil@wias-berlin.de}
\affiliation{Weierstrass Institute for Applied Analysis and Stochastics (WIAS),
Anton-Wilhelm-Amo-Stra{\ss}e 39, 10117 Berlin, Germany}

\author{Lasse Ermoneit \orcidlink{0009-0006-0329-0164}}
\affiliation{Weierstrass Institute for Applied Analysis and Stochastics (WIAS),
Anton-Wilhelm-Amo-Stra{\ss}e 39, 10117 Berlin, Germany}

\author{Lars R. Schreiber \orcidlink{0000-0003-0904-9612}}
\affiliation{RWTH Aachen University, Forschungszentrum J\"{u}lich and Institute of Quantum Information, Otto-Blumenthal-Str. 28, 52074 Aachen, Germany}
\affiliation{ARQUE Systems GmbH, Campus Boulevard 30, 52074 Aachen, Germany}

\author{Thomas Koprucki \orcidlink{0000-0001-6235-9412}}
\affiliation{Weierstrass Institute for Applied Analysis and Stochastics (WIAS),
Anton-Wilhelm-Amo-Stra{\ss}e 39, 10117 Berlin, Germany}

\author{Markus Kantner \orcidlink{0000-0003-4576-3135}}
\email{kantner@wias-berlin.de}
\affiliation{Weierstrass Institute for Applied Analysis and Stochastics (WIAS),
Anton-Wilhelm-Amo-Stra{\ss}e 39, 10117 Berlin, Germany}

\begin{abstract}
Small and device-dependent valley splittings remain a key challenge for electron spin qubits in silicon (Si), directly limiting qubit fidelity, device uniformity, and the scalability of Si-based quantum processors.
In silicon–germanium (SiGe) heterostructures, this problem can be addressed through engineering of the epitaxial layer stack.
Several heuristic strategies have been proposed to enhance the energy gap between the two nearly degenerate valley states in strained Si/SiGe quantum wells (QWs), \emph{e.g.}, sharp Si/SiGe interfaces, Ge spikes, or oscillating Ge concentrations within the QW.
Here, we develop a systematic variational optimization approach to compute optimal Ge concentration profiles that enhance selected properties of the intervalley coupling matrix element.
Our free-shape optimization framework is augmented by realistic technological constraints to ensure feasibility of the resulting epitaxial profiles and is based on an effective-mass envelope-function theory accounting for strain and compositional alloy disorder.
Previously proposed heterostructure designs are recovered as special cases of the constrained optimization problem.
Our main result is a novel heterostructure design, which we refer to as the \emph{modulated wiggle well}, providing both a large deterministic enhancement of the valley splitting and a strong suppression of disorder-induced variability.
In addition, this design enables wide electrical tunability of the valley splitting--from approximately  $200\,\upmu\mathrm{eV}$ to above $1\,\mathrm{meV}$--offering new opportunities for engineering robust and switchable silicon qubits.
\end{abstract}

\maketitle

\section{Introduction \label{sec: introduction}}

Spin qubits in silicon-germanium (SiGe) heterostructures are a promising platform for realizing fully scalable, fault-tolerant quantum computers \cite{Zwanenburg2013,Burkard2023,Hu2025}:
They offer a very small footprint, long spin coherence times enabled by nuclear spin-free isotopes \cite{Tyryshkin2011, Song2024, Daoust2026}, weak spin-orbit interaction, and compatibility with established semiconductor fabrication technology \cite{George2025,Huckemann2025,Neyens2024,Koch2025}.
Experiments have demonstrated state initialization, readout, one and two-qubit gate operations \cite{Yoneda2017, Zajac2018,Noiri2022, Xue2022, Philips2022, Mills2022, Wu2025, deFuentes2026} as well as quantum error mitigation \cite{Sohn2025} and correction \cite{Takeda2022} with high fidelity.
Scalable quantum computing architectures demand a modular processor design with coherent coupling of distant qubits to overcome crosstalk and quantum dot (QD) wiring limitations \cite{Vandersypen2017, Langrock2023, Kuenne2024, Ginzel2024}.
As a major step in this direction, coherent qubit transfer across the chip was recently demonstrated using conveyor-mode spin-qubit shuttles \cite{Seidler2022,Xue2024,Struck2024,De_Smet2025}.

\begin{figure}
\includegraphics[width=1\columnwidth]{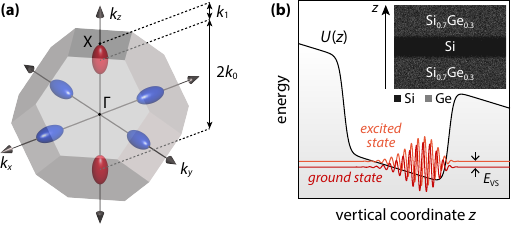}
\caption{\textbf{(a)}~Brillouin zone of the face centered cubic crystal with six valley states along the edges between the $\Gamma$ point and the six $X$ points.
In biaxially strained Si/SiGe QWs, the four valley states along the $k_{x}$ and $k_{y}$ directions (blue) are energetically separated from the two low-energy valleys along the $k_{z}$ directions (red).
\textbf{(b)}~Schematic illustration of the ground and excited valley state wave function (including the rapidly oscillating Bloch factors) in the effective confinement potential of a $\mathrm{Si}/\mathrm{Si}_{0.7}\mathrm{Ge}_{0.3}$ QW.
The energy splitting $E_{\mathrm{VS}}$ between the two valley
states is usually small in conventional QWs with smooth interfaces.
Disorder in the SiGe alloy leads to a significant volatility of the valley splitting.
The inset shows a typical realization of alloy disorder in a conventional Si/SiGe QW.
\label{fig: valley states}
}
\end{figure}

A key challenge in strained Si/SiGe quantum wells (QWs) is the existence of two nearly degenerate conduction band valley states that can lead to leakage of quantum information outside of the computational Hilbert space.
The energy splitting between the two valleys, \emph{i.e.}, the \emph{valley splitting}, see Fig.~\ref{fig: valley states}, is often small (typically $10-100\,\upmu\mathrm{eV}$) and 
highly sensitive to atomistic details of the Si/SiGe interface and alloy disorder, leading to significant device-to-device variability \cite{Klos2024, PaqueletWuetz2022, Dodson2022, DegliEsposti2024, Volmer2026}.
Presently, the valley splitting poses a major bottleneck for silicon spin qubits that critically limits the scalability of the technology platform.
Several strategies have been proposed to enhance the valley splitting,
such as engineering of sharp interfaces \cite{DegliEsposti2024}, Ge spikes \cite{McJunkin2021,Salamone2026}, superlattice barriers \cite{Zhang2013, Wang2022}, resonant Ge-layer engineering \cite{Cvitkovich2026b}, and oscillating Ge concentrations (``wiggle wells'') \cite{McJunkin2022,Feng2022,Losert2023,Gradwohl2025,Cvitkovich2026a,Rahlff2026}, often combined with shear strain engineering \cite{Woods2024,Thayil2025,Marcogliese2026}.
The inclusion of Ge, however, inevitably leads to alloy fluctuations that cause a statistical broadening of the valley splitting distribution in the device \cite{Neyens2018,Hosseinkhani2020,PaqueletWuetz2022,Pena2024,DegliEsposti2024,Klos2024,Thayil2025}.
Robust Si/SiGe qubits require a deterministic enhancement of the valley splitting that reliably exceeds the Zeeman splitting across the entire chip to avoid spin-valley hotspots \cite{Yang2013,Huang2014,Hollmann2020}.
This is especially important for shuttling-based architectures \cite{Losert2024,Volmer2024,Oda2026,Volmer2026} and large dense qubit arrays \cite{Vandersypen2017,Veldhorst2017}, where the variability of the valley splitting is probed over large domains spanning several micrometers.
Finally, a low valley splitting limits the fidelity of two-qubit gate operations (since the electrons must reside in a definite valley state) and the Pauli spin blockade readout \cite{Lai2011}.

In this work, we consider the design of the epitaxial Ge concentration profile as a constrained optimization problem to enhance the valley splitting.
Our approach is based on a multi-valley envelope function theory augmented by nonlocal empirical pseudopotential theory that accounts for the effects of strain and alloy disorder \cite{Thayil2025}.
By considering a number of different optimization objectives, we recover a variety of the heuristically known epitaxial profiles in a systematic way, including, \emph{e.g.}, narrow wells and structures with wiggle wells and Ge spikes.
Our main result is a \emph{modulated wiggle well} that outperforms the conventional sinusoidal wiggle well by enhancing the deterministic contribution to the valley splitting while simultaneously reducing the disorder-induced random component.

This paper is organized as follows: In Sec.~\ref{sec: model} we derive an expression for the intervalley coupling matrix element in Si/SiGe spin qubits, which is a brief review of the theoretical model from Ref.~\cite{Thayil2025}.
In Sec.~\ref{sec:Optimization}, we describe our variational optimization approach, including the formulation of the cost functional and details on the numerical method.
Finally, optimization results are presented and discussed in Sec.~\ref{sec:results}.
We close with a summary of results and a brief outlook in Sec.~\ref{sec: summary}.

\section{Theory of Intervalley Coupling
\label{sec: model}}

In this section, we summarize the theoretical model for the intervalley coupling matrix element employed in the optimization procedure below in Sec.~\ref{sec:Optimization}.
The model is based on multi-valley envelope function theory augmented by a statistical model to account for random alloy disorder as well as a nonlocal empirical pseudopotential model describing band structure effects in the presence of strain.
Due to the consideration of random alloy disorder, the intervalley coupling matrix element is regarded as a stochastic quantity throughout this work.
Similar effective-mass-type models were shown to be in good agreement with atomistic tight-binding models and density functional theory \cite{Losert2023, Klos2024, Cvitkovich2026a}.
Our approach enables a quantification of the disorder-induced statistical broadening of the complex-valued valley coupling matrix element and the valley splitting without any atomistic sampling of microscopic SiGe heterostructure realizations.
It is therefore computationally efficient and well suited for gradient-based optimization approaches that require fast forward evaluations and adjoint-based gradient computations.  
A comprehensive description of the model can be found in Ref.~\cite{Thayil2025}.
For a refined version with non-local potential energy operators based on exact Burt--Foreman type envelope function theory, we refer to Ref.~\cite{Ermoneit2026}.

\subsection{Hamiltonian}

We consider an electron in a gate-defined QD in a strained Si/SiGe QW grown on the {[}001{]} surface.
Biaxial strain lifts the degeneracy of the six conduction band valley states and separates the two low-energy Bloch states with wave vectors near $\mathbf{k}=\pm\mathbf{k}_{0}\approx\left(0,0,\pm0.84\right)\times2\pi/a_{0}$
from the other four conduction band valleys \cite{VandeWalle1986}, see Fig.~\ref{fig: valley states}.
The remaining two valley states are coupled via valley-orbit interaction at the heterostructure interfaces, which finally lifts their degeneracy.
The two-valley envelope function obeys the stationary, effective mass type Schrödinger equation \cite{Saraiva2009, Friesen2007, Feng2022, Thayil2025, Woods2024}
\begin{equation}
\left(\begin{array}{cc}
H_{0}\left(\mathbf{r}\right) & V_{c}\left(\mathbf{r}\right)\\
V_{c}^{*}\left(\mathbf{r}\right) & H_{0}\left(\mathbf{r}\right)
\end{array}\right)\left(\begin{array}{c}
\Psi_{+}\left(\mathbf{r}\right)\\
\Psi_{-}\left(\mathbf{r}\right)
\end{array}\right)=E\left(\begin{array}{c}
\Psi_{+}\left(\mathbf{r}\right)\\
\Psi_{-}\left(\mathbf{r}\right)
\end{array}\right),\label{eq: stationary Schroedinger}
\end{equation}
where $\Psi_{\pm}$ are the valley-specific envelope wave functions
corresponding to the Bloch waves at $\pm\mathbf{k}_{0}$. The diagonal
part of the Hamiltonian
\begin{equation}
H_{0}=-\frac{\hbar^{2}}{2m_{t}}\left(\frac{\partial^{2}}{\partial x^{2}}+\frac{\partial^{2}}{\partial y^{2}}\right)-\frac{\hbar^{2}}{2m_{l}}\frac{\partial^{2}}{\partial z^{2}}+U\left(\mathbf{r}\right)\label{eq: H0}
\end{equation}
includes the transverse and longitudinal effective masses $m_{t}$
and $m_{l}$ as well as the total confinement potential $U$. The two valley states are coupled by the interaction Hamiltonian
\begin{align}
V_{c}\left(\mathbf{r}\right) & =\mathrm{e}^{-2i\mathbf{k}_{0}\cdot\mathbf{r}}u_{+}^{*}\left(\mathbf{r}\right)u_{-}\left(\mathbf{r}\right)U\left(\mathbf{r}\right)\label{eq: Vc}\\
 & =\sum_{\mathbf{G},\mathbf{G}'}\mathrm{e}^{-i\left(\mathbf{G}-\mathbf{G}'+2\mathbf{k}_{0}\right)\cdot\mathbf{r}}c_{+}^{*}\left(\mathbf{G}\right)c_{-}\left(\mathbf{G}'\right)U\left(\mathbf{r}\right),\nonumber
\end{align}
where $\mathbf{G}$, $\mathbf{G}'$ are reciprocal lattice vectors
of the (strained) diamond crystal and $c_{\pm}\left(\mathbf{G}\right)=c_{\pm\mathbf{k}_{0}}\left(\mathbf{G}\right)$
are the plane wave expansion coefficients of the conduction band Bloch
factors $u_{\pm}\left(\mathbf{r}\right)=\sum_{\mathbf{G}}\exp{\left(i\mathbf{G}\cdot\mathbf{r}\right)}c_{\pm}\left(\mathbf{G}\right)$.

The total potential reads
\begin{equation}
U\left(\mathbf{r}\right)=U_{\mathrm{het}}\left(\mathbf{r}\right)+U_{\mathrm{QD}}\left(x,y\right)+U_{F}\left(z\right),\label{eq: total U}
\end{equation}
where $U_{\mathrm{het}}\left(\mathbf{r}\right)$ is the heterostructure potential that includes a stochastic component modeling random alloy disorder (see below).
The in-plane electrostatic confinement potential due to the gate-defined QD is given by
\begin{equation}
U_{\mathrm{QD}}\left(x,y\right)=\frac{m_{t}}{2}\left(\omega_{x}^{2}x^{2}+\omega_{y}^{2}y^{2}\right),\label{eq: U QD}
\end{equation}
where $\omega_{x}$ and $\omega_{y}$ determine the lateral extension
of the QD via $l_{i}^{2}=\hbar/\left(m_{t}\omega_{i}\right)$, $i\in\{x,y\}$.
The electric field in vertical direction contributes to the total potential as
\begin{equation}
U_{F}\left(z\right)=-e_{0}Fz,\label{eq: U F}
\end{equation}
where $e_{0}$ is the elementary charge and $F$ is the electric field strength.
Further contributions to the potential like, \emph{e.g.} electrostatic defects \cite{Ciroth2025} or charge noise are omitted in this work.

We use a nonlocal empirical pseudopotential model \cite{Rieger1993,Ungersboeck2007b}
accounting for spatially homogeneous strain to compute the effective mass tensor components $m_{t}$ and $m_{l}$, Bloch factor expansion coefficients $c_{\pm}\left(\mathbf{G}\right)$ and the valley wave vector $\mathbf{k}_{0}$ at a given strain tensor. We refer to Ref.~\cite{Thayil2025} for details. 

\subsection{Heterostructure Potential and Random Alloy Disorder \label{sec:Heterostructure-Potential}}

\begin{figure*}
\includegraphics[width=1.0\textwidth]{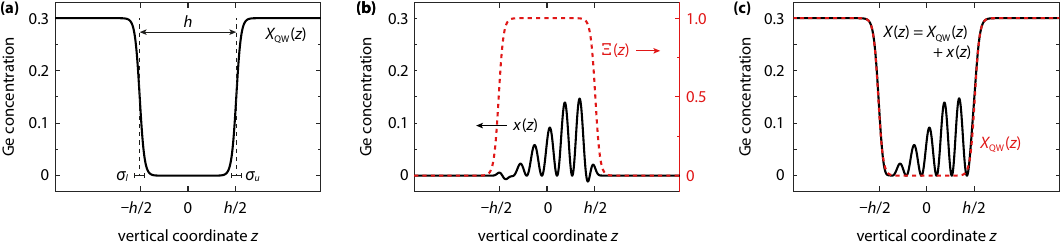}
\caption{Illustration of the decomposition of the full epitaxial profile defined in Eq.~\eqref{eq: full epitaxial profile}.
\textbf{(a)}~Fixed quantum well profile $X_{\mathrm{QW}}\left(z\right)$ as given in Eq.~\eqref{eq: QW Ge profile} with smoothed interfaces and zero Ge concentration within the well. The barrier Ge concentration is set to $X_b = 0.3$.
\textbf{(b)}~Illustration of a possible epitaxial profile modification $x\left(z\right)$ subject to the optimization problem in Sec.~\ref{sec:Optimization}. 
Via the mechanism described in Eq.~\eqref{eq: spectral constraint implementation}, the modifications are restricted to the QW domain determined by the indicator function $\Xi\left(z\right)$ given in Eq.~\eqref{eq: QW indicator}.
Near the interfaces, the plotted modification $x\left(z\right)$ is slightly negative, leading to sharper interfaces for $X\left(z\right)$ compared to  $X_\mathrm{QW}\left(z\right)$.
\textbf{(c)}~Full epitaxial profile shown as the sum of the fixed QW part and the variational modification, see Eq.~\eqref{eq: full epitaxial profile}.
In the optimization procedure, side constraints encoded in the penalty functions $J_3$ and $J_4$ set local and global restrictions on the full epitaxial profile $X\left(z\right) = X_{\mathrm{QW}}\left(z\right) + x\left(z\right)$.
The fixed QW part is shown as a red dashed line for comparison.
\label{fig: epitaxial profile opt}
}
\end{figure*}

The heterostructure confinement potential is modeled as a random field
following a scaled Bernoulli distribution
\begin{equation}
U_{\mathrm{het}}\left(\mathbf{r}\right)=\Delta E_{c}\,\Omega_{a}\sum_{i}N_{i}\delta\left(\mathbf{r}-\mathbf{R}_{i}\right)\label{eq: heterostructure U}
\end{equation}
where $N_{i}\sim\mathrm{Bernoulli}\left(p=X\left(\mathbf{R}_{i}\right)\right)$ describes the probability of finding a Ge atom at the lattice position
$\mathbf{R}_{i}$. Moreover, $\Omega_{a}$ is the atomic volume and
$\Delta E_{c}$ is the conduction band offset for Ge atoms in a Si
background lattice. We assume a layered (one-dimensional) Ge concentration
profile such that $X\left(\mathbf{r}\right)=X\left(z\right)$ depends nominally
only on the vertical coordinate. Then, the deterministic component
(\emph{i.e.}, the mean) of the heterostructure potential is directly
proportional to the Ge epitaxial profile
\begin{equation}
\left\langle U_{\mathrm{het}}\left(\mathbf{r}\right)\right\rangle =U_{\mathrm{QW}}\left(z\right)=\Delta E_{c}\,X\left(z\right).\label{eq: U_QW}
\end{equation}
The random component of the heterostructure potential
\begin{equation}
\delta U_{\mathrm{het}}\left(\mathbf{r}\right)=U_{\mathrm{het}}\left(\mathbf{r}\right)-U_{\mathrm{QW}}\left(z\right)\label{eq: random U}
\end{equation}
has zero mean $\left\langle \delta U_{\mathrm{het}}\left(\mathbf{r}\right)\right\rangle =0$
by construction and a Delta-like covariance function
\begin{align}
\langle\delta U_{\mathrm{het}}\left(\mathbf{r}\right) & \delta U_{\mathrm{het}}\left(\mathbf{r}'\right)\rangle=\label{eq: delta U covariance}\\
 & =\left(\Delta E_{c}\right)^{2}\Omega_{a}X\left(z\right)\left(1-X\left(z\right)\right)\,\delta\left(\mathbf{r}-\mathbf{r}'\right).\nonumber 
\end{align}
This corresponds to an uncorrelated substitutional alloy on lattice sites (no clustering or short-range order), leading to $\delta$‑like correlations in the continuum limit. 
Throughout this paper, we assume an epitaxial profile of the form
$X\left(z\right)=X_{\mathrm{QW}}\left(z\right)+x\left(z\right)$.
Here,
\begin{equation}
X_{\mathrm{QW}}\left(z\right)=X_{b}\left(1-\Xi\left(z\right)\right)\label{eq: QW Ge profile}
\end{equation}
is a fixed Ge concentration profile describing a smoothed step-like QW, which is supplemented by a variable modification $x\left(z\right)$ subject to a variational optimization problem described in Sec.~\ref{sec:Optimization}.
The barrier Ge concentration is fixed to $X_{b}=0.3$ and the QW indicator function describing the shape and position of the QW reads
\begin{equation}
\Xi\left(z\right) =\frac{1}{2}\left(\tanh\left(\frac{z+h/2}{\sigma_{l}}\right)-\tanh\left(\frac{z-h/2}{\sigma_{u}}\right)\right).\label{eq: QW indicator}
\end{equation}
Here, $h$ is the nominal thickness of the QW and $\sigma_{l}$ and $\sigma_{u}$
describe the width of the lower and upper interface, respectively.
The decomposition of the epitaxial profile is illustrated in Fig.~\ref{fig: epitaxial profile opt}.
Parameter values used in the numerical simulations are listed in Tab.~\ref{tab:parameters}.

\subsection{Statistical Distribution of the Intervalley Coupling Matrix Element \label{sec:Statistical-Distribution}}

Following first-order degenerate perturbation theory \cite{Friesen2007}, a perturbative expression for the intervalley coupling matrix element is obtained as \cite{Thayil2025}
\begin{align}
\Delta & =\int\mathrm{d}^{3}r\,\mathrm{e}^{-2i\mathbf{k}_{0}\cdot\mathbf{r}}u_{+}^{*}\left(\mathbf{r}\right)u_{-}\left(\mathbf{r}\right)U\left(\mathbf{r}\right)\Psi_{0}^{2}\left(\mathbf{r}\right)\label{eq: Delta}\\
 & =\sum_{\mathbf{G},\mathbf{G}'}c_{+}^{*}\left(\mathbf{G}\right)c_{-}\left(\mathbf{G}'\right)\int\mathrm{d}^{3}r\,\mathrm{e}^{-i\left(\mathbf{G}-\mathbf{G}'+2\mathbf{k}_{0}\right)\cdot\mathbf{r}}U\left(\mathbf{r}\right)\Psi_{0}^{2}\left(\mathbf{r}\right).\nonumber 
\end{align}
Here, both the intervalley coupling (\ref{eq: Vc}) and the random
alloy fluctuations (\ref{eq: random U}) were assumed to be weak. Moreover,
$\Psi_{0}\left(\mathbf{r}\right)$ is the ground state envelope wave
function satisfying the (unperturbed) single-valley eigenvalue problem
for the mean potential $\left\langle U\left(\mathbf{r}\right)\right\rangle $
(\emph{i.e.}, without random alloy fluctuations)
\begin{equation*}
\hat{H}^{\left(0\right)}\left(\mathbf{r}\right)\Psi_{0}\left(\mathbf{r}\right)=E_{0}\Psi_{0}\left(\mathbf{r}\right), 
\end{equation*}
where the single-valley Hamiltonian reads
\begin{align*}
\hat{H}^{\left(0\right)}\left(\mathbf{r}\right) & =-\frac{\hbar^{2}}{2m_{t}}\left(\frac{\partial^{2}}{\partial x^{2}}+\frac{\partial^{2}}{\partial y^{2}}\right)-\frac{\hbar^{2}}{2m_{l}}\frac{\partial^{2}}{\partial z^{2}}+\left\langle U\left(\mathbf{r}\right)\right\rangle, 
\end{align*}
and
$\left\langle U\left(\mathbf{r}\right)\right\rangle =U_{\mathrm{QW}}\left(z\right)+U_{F}\left(z\right)+U_{\mathrm{QD}}\left(x,y\right)$ is the deterministic component of the confinement potential. 
Since the Hamiltonian $\hat{H}^{(0)}$ is real-valued, we choose the envelope wave function $\Psi_0(\mathbf{r})$ real-valued throughout this work.

The intervalley coupling matrix element $\Delta$ is highly sensitive to strain. In fact, deformations of the diamond lattice structure are required to unlock an important interaction mechanism for the coupling of valley states in neighboring Brillouin zones \cite{Woods2024,Thayil2025}.
Throughout this work we assume biaxial strain $\varepsilon_{\parallel}=1.14\%$,
$\varepsilon_{\perp}=-0.88\%$ induced by the pseudomorphic lattice
matched growth of the SiGe/Si/SiGe QW \cite{Fischetti1996,Thayil2025}
and additional spatially homogeneous shear strain $\varepsilon_{s}=0.1\%$
along the {[}110{]} direction. The total strain tensor reads
\begin{align*}
\varepsilon & =\left(\begin{array}{ccc}
\varepsilon_{\parallel} & \varepsilon_{s} & 0\\
\varepsilon_{s} & \varepsilon_{\parallel} & 0\\
0 & 0 & \varepsilon_{\perp}
\end{array}\right).
\end{align*}
In the computation of the intervalley coupling strength, the strain
enters via modifications of the reciprocal lattice vectors and the
Bloch factors computed from the empirical pseudopotential model \cite{Thayil2025}.

Since the heterostructure potential (\ref{eq: heterostructure U})
includes a deterministic and a random component, the same holds true
for the intervalley coupling matrix element
\begin{equation*}
\Delta =\Delta_{\mathrm{det}}+\Delta_{\mathrm{rand}}.
\end{equation*}
The deterministic part is determined by the mean potential
\begin{equation}
\Delta_{\mathrm{det}} =\left\langle \Delta\right\rangle =\int\mathrm{d}^{3}r\,\mathrm{e}^{-2i\mathbf{k}_{0}\cdot\mathbf{r}}u_{+}^{*}\left(\mathbf{r}\right)u_{-}\left(\mathbf{r}\right)\left\langle U\left(\mathbf{r}\right)\right\rangle \Psi_{0}^{2}\left(\mathbf{r}\right),\label{eq: Delta det-1}
\end{equation}
whereas characteristic alloy disorder encoded in the stochastic potential fluctuations
(\ref{eq: random U}) governs the random component 
\begin{align*}
\Delta_{\mathrm{rand}} & =\Delta-\langle \Delta\rangle\\
 & =\int\mathrm{d}^{3}r\,\mathrm{e}^{-2i\mathbf{k}_{0}\cdot\mathbf{r}}u_{+}^{*}\left(\mathbf{r}\right)u_{-}\left(\mathbf{r}\right)\delta U_{\mathrm{het}}\left(\mathbf{r}\right)\Psi_{0}^{2}\left(\mathbf{r}\right).
\end{align*}
The value of $\Delta$ results from the summation over
a vast number of random lattice-site contributions weighted by the smooth envelope wave function.
Using the central limit theorem, it can be shown that $\Delta$ obeys (approximately)
a complex normal distribution \cite{Thayil2025}
\begin{align*}
\Delta & \sim\mathrm{ComplexNormal}\left(\Delta_{\mathrm{det}},\Gamma,C\right)
\end{align*}
with the covariance $\Gamma$ and the pseudo-covariance $C$ given as
\begin{align}
\Gamma & =\langle\left|\Delta_{\mathrm{rand}}\right|^{2}\rangle, & C & =\langle\Delta_{\mathrm{rand}}^{2}\rangle.\label{eq: Gamma and C}
\end{align}
Formulas used to evaluate the deterministic component $\Delta_{\mathrm{det}}$,
covariance $\Gamma$ and pseudo-covariance $C$ for a layered epitaxial
profile $X\left(z\right)$ are given in Appendix~\ref{sec: formulas for Delta}.
The pseudo-covariance is typically much smaller than the covariance
$\left|C\right|\ll\Gamma$, such that the distribution law of $\Delta$
is well described by the circular approximation with independent, normally distributed real and imaginary parts
\begin{align*}
\mathrm{Re}\left(\Delta\right) & \sim\mathrm{Normal}\left(\mu=\mathrm{Re}\left(\Delta_{\mathrm{det}}\right),\sigma^{2}=\frac{1}{2}\Gamma\right),\\
\mathrm{Im}\left(\Delta\right) & \sim\mathrm{Normal}\left(\mu=\mathrm{Im}\left(\Delta_{\mathrm{det}}\right),\sigma^{2}=\frac{1}{2}\Gamma\right).
\end{align*}
Using this approximation, the valley splitting
$E_{\mathrm{VS}}=2\left|\Delta\right|$ is found to obey a Rice distribution
\cite{PaqueletWuetz2022,Losert2023,Thayil2025}
\begin{equation}
E_{\mathrm{VS}} \sim\mathrm{Rice}\left(\nu=2\left|\Delta_{\mathrm{det}}\right|,\sigma^{2}=2\Gamma\right).\label{eq: Rice dist}
\end{equation}
The mean of the Rice distribution is given as \cite{Thayil2025}
\begin{equation}
\left\langle E_{\mathrm{VS}}\right\rangle   =
\sqrt{\pi\Gamma}\,f\left(\frac{\left|\Delta_{\mathrm{det}}\right|^{2}}{2\Gamma}\right), \label{eq: mean E_VS}
\end{equation}
where $f\left(x\right)=\mathrm{e}^{-x}\left(\left(1+2x\right)I_{0}\left(x\right)+2xI_{1}\left(x\right)\right)$
and $I_{\nu}\left(\cdot\right)$ denotes the modified Bessel function
of first kind.
The variance is obtained as
\begin{equation}
\mathrm{Var}\left(E_{\mathrm{VS}}\right)=4\Gamma\left(1+\frac{\left|\Delta_{\mathrm{det}}\right|^{2}}{\Gamma}\right)-\left\langle E_{\mathrm{VS}}\right\rangle^{2}.
\label{eq: variance E_VS}
\end{equation}
Finally, we introduce \emph{deterministic component
ratio}
\begin{align}
Q & =\frac{2\left|\Delta_{\mathrm{det}}\right|}{\left\langle E_{\mathrm{VS}}\right\rangle }=2\sqrt{\frac{2}{\pi}}\frac{\zeta}{f\left(\zeta^{2}\right)}\qquad\mathrm{with}\quad\zeta=\frac{\left|\Delta_{\mathrm{det}}\right|}{\sqrt{2\Gamma}},\label{eq: deterministic enhancement Q}
\end{align}
which quantifies the fraction of the expected valley splitting \eqref{eq: mean E_VS} attributable to the deterministic component of the intervalley coupling rather than disorder.
If the deterministic component dominates over the magnitude of the random component $\zeta\gg1$, the deterministic component ratio approaches unity $Q\to1$ indicating device-to-device reproducibility. On the contrary, low values of $Q$ indicate that the mean valley splitting is dominated by disorder-induced contributions.
At $\zeta\approx0.3507$, both the deterministic and the random component contribute equally to the expected valley splitting such that $Q=1/2$.

\begin{table}
\begin{tabular*}{1\columnwidth}{@{\extracolsep{\fill}}lll}
\toprule 
\textbf{Symbol} & \textbf{Description} & \textbf{Value}\tabularnewline
\midrule
$a_{0}$ & Si lattice constant & $0.543\,\mathrm{nm}$\tabularnewline
$\Delta E_{c}$ & Si/Ge conduction band offset & $0.5\,\mathrm{eV}$\tabularnewline
$\Omega_{a}$ & atomic volume & $\left(a_{0}/2\right)^{3}$\tabularnewline
$k_{0}$ & valley wave number & $0.8394\times2\pi/a_{0}$\tabularnewline
$k_{1}$ & see Eq.~(\ref{eq: k1}) & $0.1694\times2\pi/a_{0}$\tabularnewline
$m_{t}$ & transverse effective mass & $0.209\,m_{0}$\tabularnewline
$m_{l}$ & longitudinal effective mass & $0.909\,m_{0}$\tabularnewline
$h$ & thickness of the QW domain & $75\,\mathrm{ML}$\tabularnewline
$\sigma_{u}$, $\sigma_{l}$ & upper and lower interface width & $0.5\,\mathrm{nm}$\tabularnewline
$X_{b}$ & nominal Ge concentration in barrier & $0.3$\tabularnewline
$\hbar\omega_{x}$, $\hbar\omega_{y}$ & circular QD orbital energy splitting & $3\,\mathrm{meV}$\tabularnewline
$F$ & vertical electric field & $5\,\mathrm{mV/nm}$\tabularnewline
\bottomrule
\end{tabular*}

\caption{Parameters used in the numerical simulations. A complete list of parameters
including also the empirical pseudopotential model is given in Ref.~\cite{Thayil2025}.
Here, $m_{0}$ is the vacuum electron mass and $\mathrm{ML}=a_{0}/4$
is the silicon monolayer thickness.}

\label{tab:parameters}
\end{table}

\section{Epitaxial Profile Optimization \label{sec:Optimization}}

In Si/SiGe heterostructures, the intervalley coupling matrix element can be tailored by the design of the epitaxial layer stack.
Different heuristics for enhancement of the magnitude of the valley splitting were recently discussed in the literature, including 
wiggle wells \cite{McJunkin2022,Feng2022,Losert2023,Woods2024,Thayil2025},
sharp interfaces \cite{DegliEsposti2024},
Ge spikes \cite{McJunkin2021,Losert2023},
narrow wells \cite{Losert2023} and
uniform Ge concentrations in the QW \cite{PaqueletWuetz2022,Losert2023,Losert2024}.
All these Ge concentration profiles affect both the deterministic and the random component of the valley coupling matrix element in a specific way.

In this section, we address the design of the epitaxial Ge concentration profile as a variational optimization problem to compute optimized epitaxial
profiles that enhance selected properties of the intervalley coupling matrix element.
To this end, we consider a Ge concentration profile
of the form 
\begin{equation}
X\left(z\right)=X_{\mathrm{QW}}\left(z\right)+x\left(z\right),   
\label{eq: full epitaxial profile}
\end{equation}
where $X_{\mathrm{QW}}\left(z\right)$ is a fixed QW profile
(smoothed step function, see Sec. \ref{sec:Heterostructure-Potential})
and $x\left(z\right)$ is a modification subject to 
optimization.
The epitaxial profile modification implies a corresponding modification of the potential energy according to Eq.~(\ref{eq: heterostructure U}).

\subsection{Optimization Objectives \label{sec: objectives}}

We consider a number of different optimization objectives.
In each case, we seek to minimize a non-negative cost functional $J_0$ involving combinations of the deterministic contribution to the valley splitting
$\nu\left(x,\psi_{0}\right)=2\left|\Delta_{\mathrm{det}}\left(x,\psi_{0}\right)\right|$ and the disorder-induced contribution quantified by the covariance $\Gamma\left(x,\psi_{0}\right)$.
All numerical computations are based on the separation ansatz described in Appendix~\ref{sec: formulas for Delta}, which reduces the problem to an effective one-dimensional Schrödinger problem depending only on the vertical component  $\psi_{0}\left(z\right)$ of the envelope wave function determined by the Ge concentration profile $x\left(z\right)$.

The different objectives under consideration read:
\begin{enumerate}[label=(\Alph*)]
\item  Maximization of the deterministic component
\begin{equation*}
J_{0}^{\left(A\right)}\left(x,\psi_{0}\right)=\frac{E_{\mathrm{ref}}}{\nu\left(x,\psi_{0}\right)},
\end{equation*}
where the deterministic contribution to the valley splitting is $\nu\left(x,\psi_{0}\right)=2\left|\Delta_{\mathrm{det}}\left(x,\psi_{0}\right)\right|$.
\item Minimization of the ratio between disorder-induced and deterministic components
\begin{equation*}
J_{0}^{\left(B\right)}\left(x,\psi_{0}\right)=\frac{\sqrt{2\Gamma\left(x,\psi_{0}\right)}}{\nu\left(x,\psi_{0}\right)}=\frac{1}{2\zeta\left(x,\psi_{0}\right)}
\end{equation*}
to maximize the \emph{reliable enhancement}. As the deterministic
component ratio $Q$ is a monotonic function of $\zeta$, see Eq.~(\ref{eq: deterministic enhancement Q}),
this objective simultaneously covers the maximization of $Q$.
\item Minimization of the disorder-induced random component
\begin{equation*}
J_{0}^{\left(C\right)}\left(x,\psi_{0}\right)=\frac{\sqrt{2\Gamma\left(x,\psi_{0}\right)}}{E_{\mathrm{ref}}}
\end{equation*}
for reduction of the valley splitting volatility.
\end{enumerate}
Some of the cost functionals include an arbitrary reference energy
set to $E_{\mathrm{ref}}=1\,\mathrm{meV}$ for non-dimensionalization.
The optimization objectives $J_{0}$ are supplemented by further constraints, leading to the total cost functional described in the following section.

\subsection{Cost Functional}
We consider a cost functional for minimization consisting of a number of additive components capturing the optimization objective and constraints.
The total cost functional reads
\begin{align}
J\left(x,\psi_{0},E_{0},\chi,\kappa\right) & =J_{0}\left(x,\psi_{0}\right)+J_{1}\left(x,\psi_{0},\chi,E_{0}\right)\label{eq: cost functional}\\
 & \phantom{=}+J_{2}\left(\kappa,\psi_{0}\right)+J_{3}\left(x\right)+J_{4}\left(x\right),\nonumber 
\end{align}
where $J_{0}\left(x,\psi_{0}\right)$ is the figure of merit (see Sec.~\ref{sec: objectives} above) that captures the key objective for optimization, depending on the Ge epitaxial profile $x$ and the ground state envelope wave function $\psi_{0}$.

The first constraint ensures that the envelope wave function $\psi_{0}$ is an eigenfunction of the Schrödinger eigenvalue problem
\begin{equation*}
J_{1}\left(x,\psi_{0},\chi,E_{0}\right) =\int\mathrm{d}z\,\chi\left(z\right)\left(\hat{H}^{\left(0\right)}-E_{0}\right)\psi_{0}\left(z\right),
\end{equation*}
where the deterministic, single-valley Hamiltonian for the longitudinal
problem includes the unknown Ge epitaxial profile $x\left(z\right)$
as a variable contribution to the potential energy
\begin{equation}
\hat{H}^{\left(0\right)} =-\frac{\hbar^{2}}{2m_{l}}\frac{\partial^{2}}{\partial z^{2}}+\Delta E_{c}\left(X_{\mathrm{QW}}\left(z\right)+x\left(z\right)\right)+U_{F}\left(z\right).
\label{eq: single-valley Hamiltonian}
\end{equation}
The ground state wave function $\psi_{0}$ and energy level $E_{0}$
are selected via suitable heuristics
\footnote{In the presence of electric fields, we always select the lowest energy bound state that has a strong overlap with the (tilted) QW domain.}.
As the Hamiltonian \eqref{eq: single-valley Hamiltonian} is real, we assume real-valued eigenfunctions $\psi_0$ throughout this paper.
Finally, $\chi\left(z\right)$ is a (real-valued) adjoint wave function that plays the role of a space-dependent Lagrange multiplier.

The next term in the cost functional
\begin{equation*}
J_{2}\left(\kappa,\psi_{0}\right) =\kappa\left(1-\int\mathrm{d}z\,\psi_{0}^{2}\left(z\right)\right)
\end{equation*}
enforces normalization of the wave function using a Lagrange multiplier $\kappa$.
This constraint is indispensable to ensure uniqueness of the optimum and solvability of the adjoint state equation, see Sec.~\ref{sec:adjoint state equation} below.

The admissible range for the local Ge concentration must be limited by lower and upper bounds $x_{\mathrm{min}}\leq X\left(z\right)\leq x_{\mathrm{max}}$, where the total epitaxial profile $X\left(z\right)=X_{\mathrm{QW}}\left(z\right)+x\left(z\right)$ consists of the fixed QW profile $X_{\mathrm{QW}}\left(z\right)$ and the modification $x\left(z\right)$, which is subject to the free-shape optimization problem.
We choose $x_{\mathrm{min}}=0$ and $x_{\mathrm{max}}=X_{b}$.
This requirement is included in the cost functional via the penalty term
\begin{equation*}
    J_{3}\left(x\right)=\frac{\alpha}{h}\int\mathrm{d}z\,g\left(X_{\mathrm{QW}}\left(z\right)+x\left(z\right)\right),
\end{equation*}
where the function
\begin{align*}
g\left(x\right) & =\begin{cases}
x_{\mathrm{min}}-x & x<x_{\mathrm{min}},\\
0 & x_{\mathrm{min}}\leq x\leq x_{\mathrm{max}},\\
x-x_{\mathrm{max}} & x>x_{\mathrm{max}}.
\end{cases}
\end{align*}
penalizes values of $x$ outside of the admissible range. The parameter
$\alpha$ is a penalty value (taken to be large).
If the epitaxial profile is within the admissible range, the contribution to the cost functional vanishes.
The quantum well thickness $h$ appears here to nondimensionalize the cost functional.

Finally, the overall Ge content in the QW is fixed at a target Ge concentration $x_\mathrm{Ge}$, which is enforced by a second penalty term
\begin{equation*}
J_{4}\left(x\right)=\frac{\beta}{2}\left(\overline{X}\left(x\right)-x_{\mathrm{Ge}}\right)^{2}
\end{equation*}
with another penalty value $\beta$. Here, $\overline{X}$ quantifies the mean Ge concentration in the QW
\begin{align}
\overline{X}\left(x\right) & =\frac{1}{h}\int\mathrm{d}z\,\Xi\left(z\right)\left(X_{\mathrm{QW}}\left(z\right)+x\left(z\right)\right). \label{eq: mean Ge content}
\end{align}
The main reason to restrict the total Ge amount is to avoid strong enhancements of spin-orbit coupling strength in the QW at too high Ge concentrations \cite{Woods2023}.
Without this restriction, the algorithm usually tends to introduce large amounts of Ge, because the detrimental enhancement of spin-orbit coupling is not included in the present model.
In the optimization procedure, the admissible amount $x_{\mathrm{Ge}}$ of Ge must be distributed within the QW to minimize the overall cost functional while respecting all constraints.

\subsection{Variational Minimization and Optimality Conditions}

We seek a minimum of the cost functional by considering a small
variation of the epitaxial profile $x\left(z\right)\to x\left(z\right)+\delta x\left(z\right)$,
which induces a variation of the ground state envelope wave function
$\psi_{0}\to\psi_{0}+\delta\psi_{0}$, the ground state energy level
$E_{0}\to E_{0}+\delta E_{0}$ and the Lagrange multipliers $\chi\left(z\right)\to\chi\left(z\right)+\delta\chi\left(z\right)$
and $\kappa\to\kappa+\delta\kappa$.
The condition for the minimum
\begin{align*}
0 &\stackrel{!}{=} \delta J\left(x,\psi_{0},E_{0},\chi,\kappa\right)\\
 & =J\left(x+\delta x,\psi_{0}+\delta\psi_{0},E_{0}+\delta E_{0},\chi+\delta\chi,\kappa+\delta\kappa\right)\\
 & \phantom{=}-J\left(x,\psi_{0},E_{0},\chi,\kappa\right)
\end{align*}
leads to
\begin{align*}
\delta J & =\int\mathrm{d}z\,\bigg[\frac{\delta J_{0}\left(x,\psi_{0}\right)}{\delta x}+\Delta E_{c}\chi\left(z\right)\psi_{0}\left(z\right)+\frac{\alpha}{h}g'\left(X\left(z\right)\right)\\
 & \hphantom{=\int\mathrm{d}z\,\bigg(}+\frac{\beta}{h}\left(\overline{X}\left(x\left(z\right)\right)-x_{\mathrm{Ge}}\right)\Xi\left(z\right)\bigg]\,\delta x\left(z\right)\\
 & \phantom{=}+\int\mathrm{d}z\,\bigg[\frac{\delta J_{0}\left(x,\psi_{0}\right)}{\delta\psi_{0}}-2\kappa\psi_{0}\left(z\right)\\
 & \phantom{=+\int\mathrm{d}z\,\bigg[}+\left(\hat{H}^{\left(0\right)}-E_{0}\right)\chi\left(z\right)\bigg]\,\delta\psi_{0}\left(z\right)\\
 & \phantom{=}+\int\mathrm{d}z\,\delta\chi\left(z\right)\left(\hat{H}^{\left(0\right)}-E_{0}\right)\psi_{0}\left(z\right)\\
 & \phantom{=}-\delta E_{0}\int\mathrm{d}z\,\chi\left(z\right)\psi_{0}\left(z\right)+\delta\kappa\left(1-\int\mathrm{d}z\,\psi_{0}^{2}\left(z\right)\right).
\end{align*}
From the optimality condition $\delta J=0$, we extract a number of
necessary conditions:\begin{subequations}\label{eq: optimality conditions}
\begin{enumerate}[label=(\roman*)]
\item  Stationary Schrödinger equation (since  $\delta\chi$
is free): The ground state wave function and energy level must be
an eigenstate of the single-valley Hamiltonian (\ref{eq: single-valley Hamiltonian})
including the adapted epitaxial profile
\begin{equation}
\hat{H}^{\left(0\right)}\psi_{0}\left(z\right)  = E_{0}\psi_{0}\left(z\right).\label{eq: stationary Schroedinger equation (optimality)}
\end{equation}
\item Normalization (since $\delta\kappa$ is free): The ground
state wave function must be normalized over the full domain
\begin{equation}
\int\mathrm{d}z\,\psi_{0}^{2}\left(z\right)  =1.\label{eq: normalization condition}
\end{equation}
\item Adjoint state equation (since $\delta\psi_{0}$ is free):
The adjoint solves the linear Schrödinger-type equation with inhomogeneous
right hand side depending on the ground state wave function
\begin{equation}
\left(\hat{H}^{\left(0\right)}-E_{0}\right)\chi\left(z\right) =2\kappa\psi_{0}\left(z\right)-\frac{\delta J_{0}\left(x,\psi_{0}\right)}{\delta\psi_{0}}.\label{eq: adjoint state equation}
\end{equation}
\item Orthogonality condition (since $\delta E_{0}$ is free):
The adjoint wave function must be orthogonal to the
ground state envelope wave function
\begin{equation}
\int\mathrm{d}z\,\chi\left(z\right)\psi_{0}\left(z\right) =0.\label{eq: orthogonality condition}
\end{equation}
\end{enumerate}
\end{subequations}
When the conditions (i)--(iv) are satisfied, the
gradient of the cost functional with respect to the epitaxial profile
is given as
\begin{align}
\frac{\delta J\left(x,\psi_{0}\right)}{\delta x} & =\frac{\delta J_{0}\left(x,\psi_{0}\right)}{\delta x}+\Delta E_{c}\chi\left(z\right)\psi_{0}\left(z\right)\label{eq: gradient delta J/delta x}\\
 & \phantom{=}+\frac{\alpha}{h}g'\left(X_{\mathrm{QW}}\left(z\right)+x\left(z\right)\right)\nonumber \\
 & \phantom{=}+\frac{\beta}{h}\left(\overline{X}\left(x\left(z\right)\right)-x_{\mathrm{Ge}}\right)\Xi\left(z\right).\nonumber 
\end{align}
The gradient is employed to iteratively update the epitaxial profile modification $x\left(z\right)$ until
convergence, see Sec.~\ref{sec: numerical method}. Next to the direct derivative of the key objective functional
$J_{0}$ with respect to the Ge concentration profile, the gradient
involves corrections due to the side constraints $J_{1}$ (eigenstate
constraint), $J_{3}$ (admissible Ge range) and $J_{4}$ (Ge budget). 

\subsection{Solution of the Adjoint State Equation \label{sec:adjoint state equation}}

The operator on the left hand side of the adjoint state equation (\ref{eq: adjoint state equation})
is rank deficient since $E_{0}$ is an eigenvalue of $\hat{H}^{\left(0\right)}$.
Thus, the operator $\big(\hat{H}^{\left(0\right)}-E_{0}\big)$
is not invertible (zero eigenvalue) and care must be taken when solving (\ref{eq: adjoint state equation})
for the adjoint wave function $\chi\left(z\right)$. We emphasize
that the latter is necessary to compute the gradient (\ref{eq: gradient delta J/delta x})
driving the minimization procedure.

The problem (\ref{eq: adjoint state equation}) is solvable, if the
right hand side has no component along the eigenspace of $E_{0}$
(Fredholm alternative). By projecting Eq.~(\ref{eq: adjoint state equation})
on the ground state envelope wave function (from the left), we obtain
by using Eqs.~(\ref{eq: stationary Schroedinger equation (optimality)})
and (\ref{eq: normalization condition}) the condition
\begin{align*}
 & \underbrace{\int\mathrm{d}z\,\psi_{0}\left(z\right)\left(\hat{H}^{\left(0\right)}-E_{0}\right)\chi\left(z\right)}_{=0}=\\
 & \qquad=2\kappa\underbrace{\int\mathrm{d}z\,\psi_{0}^{2}\left(z\right)}_{=1}-\int\mathrm{d}z\,\psi_{0}\left(z\right)\frac{\delta J_{0}\left(x,\psi_{0}\right)}{\delta\psi_{0}}.
\end{align*}
Solving this for the Lagrange multiplier $\kappa$ yields
\begin{equation}
\kappa =\frac{1}{2}\int\mathrm{d}z\,\psi_{0}\left(z\right)\frac{\delta J_{0}\left(x,\psi_{0}\right)}{\delta\psi_{0}},\label{eq: kappa}
\end{equation}
which ensures feasibility (but not yet uniqueness) of the solution.
Substituting (\ref{eq: kappa}) in (\ref{eq: adjoint state equation}),
the adjoint state equation becomes
\begin{align}
\left(\hat{H}^{\left(0\right)}-E_{0}\right)\chi & \left(z\right)=R\left(z\right),\label{eq: adjoint problem with modified rhs}
\end{align}
where the modified right hand side
\begin{align*}
R\left(z\right) & =\psi_{0}\left(z\right)\int\mathrm{d}z'\,\psi_{0}\left(z'\right)\frac{\delta J_{0}\left(x\left(z'\right),\psi_{0}\left(z'\right)\right)}{\delta\psi_{0}}\\
 & \phantom{=}-\frac{\delta J_{0}\left(x\left(z\right),\psi_{0}\left(z\right)\right)}{\delta\psi_{0}}.
\end{align*}
is orthogonal to the envelope wave function by construction:
\begin{equation*}
\int\mathrm{d}z\,\psi_{0}\left(z\right)R\left(z\right)  =0.
\end{equation*}
Finally, an explicit solution of Eq.~\eqref{eq: adjoint problem with modified rhs} is constructed by expanding the adjoint
wave equation in the complete, orthonormal eigenbasis $\left\{ \psi_{n},E_{n}\right\} $
of $\hat{H}^{\left(0\right)}$
\begin{equation*}
\chi\left(z\right) =\sum_{n}a_{n}\psi_{n}\left(z\right),
\end{equation*}
which allows to formulate the solution via a Green's function 
\begin{equation}    
\chi\left(z\right)  =\int\mathrm{d}z'\,G\left(z,z'\right)R\left(z'\right).\label{eq: formal solution chi}
\end{equation}
The expansion coefficients $a_{n}$ for $n\neq0$ are uniquely determined by
Eq.~(\ref{eq: adjoint problem with modified rhs})
and obtained as
\begin{equation*}
a_{n} =\frac{1}{E_{n}-E_{0}}\int\mathrm{d}z\,\psi_{n}\left(z\right)R\left(z\right)\qquad\left(n\neq0\right).
\end{equation*}
The missing component $a_{0}$ is fixed by the orthogonality condition
(\ref{eq: orthogonality condition}) and obtained as $a_{0}=0$. Finally, this yields
\begin{equation}
G\left(z,z'\right) =\sum_{n\neq0}\frac{\psi_{n}\left(z\right)\psi_{n}\left(z'\right)}{E_{n}-E_{0}}.\label{eq: Green's function}
\end{equation}
Thus, accurate computation of the gradient (\ref{eq: gradient delta J/delta x}) is computationally expensive, as it involves complete diagonalization of the single-valley Hamiltonian (\ref{eq: single-valley Hamiltonian}).

\subsection{Spectral Constraint \label{sec:Spectral-Constraint}}

Without further constraints on the epitaxial profile, the optimization algorithm will in most cases converge to rapidly oscillating profiles dominated by a Fourier component at wave number $2k_{0}$ (short-period wiggle-well).
This period boosts the direct coupling of the two valley states within the same Brillouin zone \cite{Feng2022,Thayil2025, Cvitkovich2026a} and thus leads to large enhancements of the valley splitting.
This high modulation frequency of the Ge concentration is very close to the atomic monolayer thickness and therefore hard to realize in practice.
In order to focus on epitaxially feasible heterostructures \cite{Gradwohl2025},
we need to rule out these rapidly modulated solutions. 
To this end, we impose a spectral constraint blocking out high frequency components by replacing
\begin{equation}
x\left(z\right)  \to\tilde{x}\left(z\right)=\Xi\left(z\right)x_{K}\left(z\right),\label{eq: spectral constraint implementation}
\end{equation}
in the cost functional, where $\Xi\left(z\right)$ is the indicator function, see Eq.~\eqref{eq: QW indicator}, and
\begin{align*}
x_{K}\left(z\right) & =\left(K*x\right)\left(z\right)=\int\mathrm{d}z'\,K\left(z-z'\right)x\left(z'\right)\\
 & =\int\frac{\mathrm{d}k}{2\pi}\,\mathrm{e}^{-ikz}K\left(k\right)x\left(k\right)
\end{align*}
is a convolution with a filter function $K$. Throughout this
work, we choose a rectangular low-pass filter with Fourier
representation
\begin{equation*}
K\left(k\right) =\Theta\left(k_{c}-\left|k\right|\right),
\end{equation*}
where $k_{c}$ is the cutoff wave number and $\Theta$ is the Heaviside
step function. The filtered profile is multiplied with
the QW indicator function \eqref{eq: QW indicator} to suppress oscillations (ringing) outside of the QW domain.
The last step potentially reintroduces short wavelength components to the Ge concentration profile, but only at a significantly lower level.
For slowly changing Ge concentration
profiles localized within the QW domain, the substitution (\ref{eq: spectral constraint implementation})
keeps the input profile practically invariant. 

By carrying out this replacement directly on the level of the cost
functional
\begin{equation*}
J\left(x\right)\to J\left(\tilde{x}\left(z\right)\right)=J\left(\Xi\left(z\right)\left(K*x\right)\left(z\right)\right),
\end{equation*}
the corresponding gradient is obtained as
\begin{align}
\frac{\delta J\left(\tilde{x}\left(z\right)\right)}{\delta\tilde{x}\left(z\right)}
&=
\left(K*\left[\Xi\,\frac{\delta J\left(x\right)}{\delta x}\right]\right)\left(z\right)\label{eq: gradient with spectral constraint}\\
 & =\int\mathrm{d}z'\,K\left(z-z'\right)\Xi\left(z'\right)\frac{\delta J\left(x\left(z'\right)\right)}{\delta x\left(z'\right)}.\nonumber 
\end{align}
Hence, the gradient of the modified functional is obtained from the
gradient of the original functional after multiplication with the indicator function $\Xi$ and subsequent low-pass filtration, \emph{i.e.}, opposite ordering compared to Eq.~(\ref{eq: spectral constraint implementation}).
We point out that simple low-pass filtering of the gradient alone
(which is a frequently chosen heuristic to block out undesired frequency components in optimization problems \cite{Werschnik2007, Gross1992}) without
corresponding modification of the cost functional will lead to a mismatch of function and gradient.
This can drastically degrade or even inhibit the convergence of the gradient-based minimization method used here.
Throughout this work, the substitutions
(\ref{eq: spectral constraint implementation}) and (\ref{eq: gradient with spectral constraint})
are tacitly employed everywhere and not explicitly indicated by notation. 

\subsection{Numerical Optimization Method \label{sec: numerical method}}

Since the ground state wave function $\psi_{0}\left(z\right)=\psi_{0}\left(x\left(z\right)\right)$
is uniquely determined by the epitaxial profile $x\left(z\right)$
via Eq.~(\ref{eq: stationary Schroedinger equation (optimality)}), the cost
functional can be written as a functional of the Ge concentration
profile only $J\left(x,\psi_{0}\right)\to J\left(x\right)$.
To linear order, an update $\Delta x=\Delta x\left(z\right)$ of the
epitaxial profile leads to
\begin{equation*}
J\left(x+\Delta x\right)  =J\left(x\right)+\int\mathrm{d}z\,\frac{\delta J\left(x\left(z\right)\right)}{\delta x}\Delta x\left(z\right)+O\left(\Delta x^{2}\right).
\end{equation*}
In order to obtain a reduction of the cost functional $J\left(x+\Delta x\right)<J\left(x\right)$,
the update is chosen as
\begin{equation*}
\Delta x\left(z\right) =-\int\mathrm{d}z'\,B\left(z,z'\right)\frac{\delta J\left(x\left(z'\right)\right)}{\delta x},
\end{equation*}
where the functional derivative is computed according to Eq.~(\ref{eq: gradient delta J/delta x})
and $B\left(z,z'\right)$ positive definite.

We employ the L-BFGS method with a Wolfe line search \cite{Nocedal2006} for numerical optimization of the cost functional (\ref{eq: cost functional}).
The method iteratively constructs low-rank approximations of the inverse Hessian $B$ to achieve superlinear convergence.
The optimization procedure is initiated from a flat profile $x\left(z\right)=0$ (no Ge).
The method is restarted multiple times after every 100 iterations (memory reset), while incrementally increasing the penalty value $\beta$ from $10^{4}$ to $10^{6}$.
This strategy provides flexibility in variation of the Ge profile in early stages of the procedure (with relaxed Ge budget constraint at low $\beta$), whereas the constraint is met with high accuracy at the end of the procedure (large $\beta$).
The penalty value to enforce the admissible Ge range is fixed to $\alpha=10^{5}$.
We use a limited memory, storing only the 20 most recent gradients to approximate the quasi-Hessian inverse in each step.
The Schrödinger equation \eqref{eq: stationary Schroedinger equation (optimality)} is solved using a finite difference method with periodic boundary conditions.
The domain is chosen large enough that the envelope wave function is negligible at the boundaries and results are insensitive to the domain size.

\section{Optimization Results \label{sec:results}}

\begin{figure*}
\includegraphics[width=0.94\textwidth]{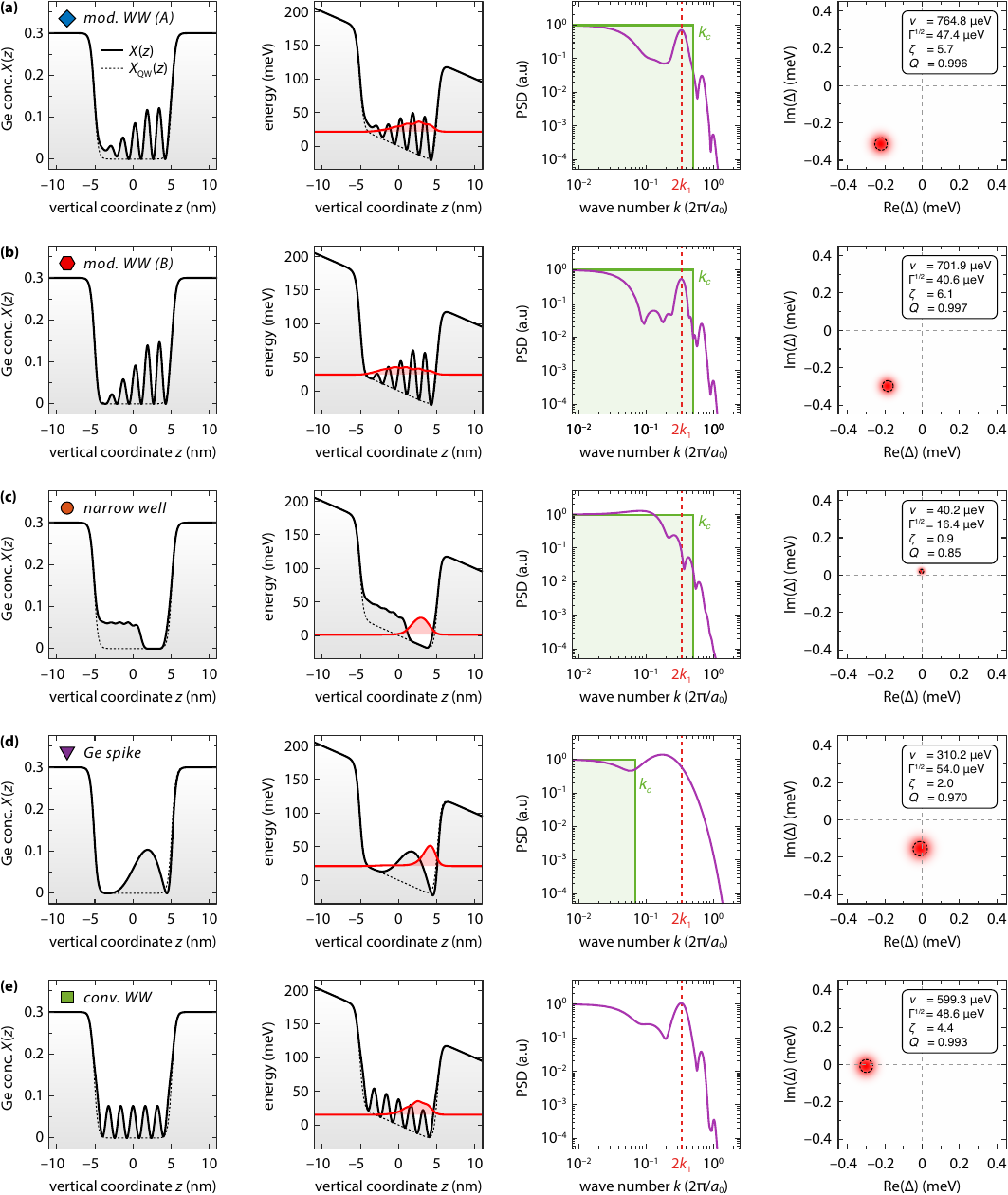}\caption{Optimization results for a fixed Ge budget $x_{\mathrm{Ge}}=5\%$ distributed within the QW.
\textbf{(a)}~Optimization with the cost functional $J_{0}^{\left(A\right)}$ (maximum deterministic component) at a cutoff wave number $k_{c}=0.5\times2\pi/a_{0}$ yields a modulated wiggle well.
\textbf{(b)}~A similar structure is found when optimizing with the cost functional $J_{0}^{\left(B\right)}$ (reliable enhancement) at the same cutoff wave number.
\textbf{(c)}~Optimization with the cost functional $J_{0}^{\left(C\right)}$ (minimum random component) at the same $k_{c}$ yields a narrow well.
\textbf{(d)}~Epitaxial profile with a Ge spike obtained by maximizing the deterministic component (with cost functional $J_{0}^{\left(A\right)}$ ) at a reduced cutoff wave number $k_{c}=0.07\times2\pi/a_{0}$.
\textbf{(e)}~Conventional wiggle well with wave number $2k_{1}$ for comparison.
In all plots, the first column shows the epitaxial
Ge profile and the second column shows the potential energy and ground-state envelope wave function at $F=5\,\mathrm{mV/nm}$.
The third column shows the power spectral density (PSD) of the function $S\left(z\right)$, see Eq.~\eqref{eq: function S}, and the fourth column illustrates the statistical distribution of the intervalley coupling matrix element $\Delta$ in the complex plane.
\label{fig: optim results}
}
\end{figure*}

The results of the numerical optimization of cost functionals with
objectives (A)--(C) are shown in Fig.~\ref{fig: optim results}\,(a)--(d)
and discussed in the following. Parameters used are listed in Tab.~\ref{tab:parameters}.

\subsection{Modulated Wiggle Well}

When optimizing for the maximization of the deterministic component (A) or reliable enhancement (B), the numerical procedure converges to an oscillating Ge concentration profile with a dominant wave number near $2k_{1}$, where 
\begin{equation}
k_{1}=\frac{2\pi}{a_{0}}\left(1-\varepsilon_{\perp}\right)-k_{0}\label{eq: k1}
\end{equation}
is the reciprocal space distance from the valley minimum to the (strained) Brillouin zone edge.
This corresponds to a wavelength of $\lambda=\pi/k_{1}\approx1.6\,\mathrm{nm}\approx 11.8\,\mathrm{ML}$, see Fig.~\ref{fig: optim results}\,(a)--(b).
This type of Ge concentration profile is closely related to the \emph{long-period wiggle well}, which will be denoted as \emph{conventional wiggle well} in the following, that has been extensively discussed in the literature \cite{McJunkin2022, Feng2022, Losert2023, Woods2023, Woods2024, Thayil2025, Gradwohl2025}.
The periodicity of the heterostructure potential boosts the coupling
of the valley states in neighboring Brillouin zones separated by $2k_{1}$
in reciprocal space, which is in resonance with the epitaxial profile
modulation. The conventional wiggle well with sinusoidal Ge concentration
profile is shown in Fig.~\ref{fig: optim results}\,(e) for reference.

\begin{figure}
\includegraphics[width=1\columnwidth]{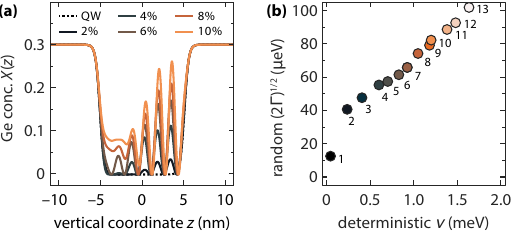}

\caption{Modulated wiggle well optimizing the reliable enhancement (B)
for different total Ge concentrations $x_{\mathrm{Ge}}$ in the QW. \textbf{(a)}~Optimized
epitaxial profiles for different Ge budgets $x_{\mathrm{Ge}}=\left\{ 2\%,4\%,6\%,8\%,10\%\right\} $.
\textbf{(b)} Deterministic $\nu$ and disorder-induced contributions
$\sqrt{2\Gamma}$ to the intervalley coupling matrix element for the modulated wiggle wells optimized at different Ge concentrations.
The labels
indicate the respective Ge concentration in percent.
\label{fig:mod_WW different Ge}
}
\end{figure}

The epitaxial profiles that satisfy our optimization objectives (A)
or (B) are referred to as \emph{modulated wiggle wells}.
These structures differ from the conventional wiggle well by a modulation of the Ge concentration amplitude, which is optimized for a particular electric field in the growth direction.
The envelope wave function of the ground state electron in these structures is stretched out over the entire QW domain in order to enhance the overlap of the wave function with the Ge modulations.
In this way, the resonance in the intervalley coupling
matrix element is boosted in an optimal way, leading to a strong enhancement of the deterministic component $\nu$.
This is reflected by the strong peak in the power spectral density (PSD) of the product
\begin{equation}
    S\left(z\right)=\left(\Delta E_{c}\,X\left(z\right)+U_{F}\left(z\right)\right)\psi_{0}^{2}\left(z\right),
    \label{eq: function S}
\end{equation}
which enters Eq.~(\ref{eq: Delta det (only z)}) and is shown in the third column of Fig.~\ref{fig: optim results}\,(a)--(b).
Both modulated wiggle wells are very similar, although different optimization
objectives are pursued. In fact, the modulated wiggle well (A) has
a slightly larger deterministic component $\nu$ than the modulated
wiggle well (B), which can be seen from the larger distance of the
center of the probability density of the intervalley coupling matrix element
$\Delta$ from the origin of the complex plane, see fourth column
of Fig.~\ref{fig: optim results}\,(a)--(b). In turn, the modulated
wiggle well (B) provides an enhanced ratio of (large) deterministic
and (small) disorder-induced contributions, simultaneously. Both structures
outperform the conventional wiggle well at the same Ge content.

We have repeated the optimization for the objective (B) for different
total Ge budgets in the QW domain. The results shown in Fig.~\ref{fig:mod_WW different Ge}\,(a)
are variants of the modulated wiggle well with increasing Ge amplitudes
for large Ge budgets, in particular close to the upper interface.
All modulated wiggle wells computed via the optimization procedure
show a monotonic increase of both the deterministic and disorder-induced
contributions to the intervalley coupling matrix element for growing Ge
content as shown in Fig.~\ref{fig:mod_WW different Ge}\,(b). We
emphasize that higher Ge concentrations also increase the spin-orbit
interaction strength \cite{Woods2023}, which has not been taken
into account in the present study.

\begin{figure}
\includegraphics[width=1\columnwidth]{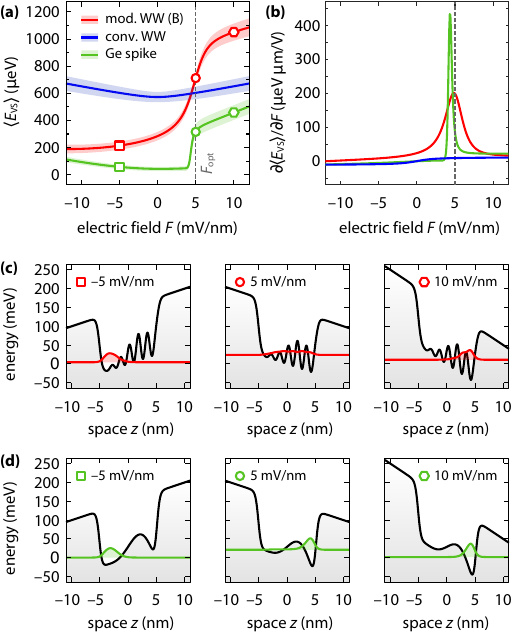}
\caption{\textbf{(a)}~Mean valley splitting $\langle E_{\mathrm{VS}}\rangle$ of the modulated wiggle well (red), the Ge spike (green) and the conventional long-period wiggle well (blue) as a function of the applied electric field $F$.
All structures contain in total $5\%$ Ge in the QW domain.
Due to the asymmetry in their epitaxial profiles, the modulated wiggle well and the Ge spike exhibit a strong, non-symmetric field dependency, which allows for tuning into a high and low valley splitting regime. 
For comparison, the field dependency of the conventional wiggle well
is only weak and fully symmetric under inversion of the field strength $F\leftrightarrow-F$.
The shaded regions indicate the {[}25\%~,75\%{]} percentile interval of the Rice distribution (\ref{eq: Rice dist}).
\textbf{(b)}~Sensitivity of the mean valley splitting to the electric field $\partial\langle E_{\mathrm{VS}}\rangle/\partial F$
as a function of the applied electric field.
The field sensitivity of the two non-symmetric structures is peaked around the design field strength $F_{\mathrm{opt}}$.
In contrast, the conventional wiggle well shows only a weak field dependency.
\textbf{(c)}~Cross section of the total potential energy landscape and the electronic ground state envelope wave function for the modulated wiggle well at different electric fields.
\textbf{(d)}~Same cross sections for the Ge spike.
The localization of the envelope wave function is strongly dependent on the electric field such that different parts of the non-symmetric epitaxial profiles are probed in the low and high valley splitting regimes.
\label{fig:mod_WW-electric field dep}
}
\end{figure}

The modulated wiggle well offers a wide tunability range of the
valley splitting via the vertical electric field. Figure~\ref{fig:mod_WW-electric field dep}\,(a) shows the mean valley splitting $\langle E_{\mathrm{VS}}\rangle$ for electrons in a modulated wiggle well (B) with $x_{\mathrm{Ge}}=5\%$
optimized at $F=F_\mathrm{opt}=5\,\mathrm{mV/nm}$ and a conventional $2k_{1}$ wiggle well with the same overall Ge content as a function of the applied electric field $F$.
While the mean valley splitting in a conventional wiggle well is only weakly dependent on the electric field, the opposite
is true for the modulated wiggle well.
We observe a wide range of tunability from $\langle E_{\mathrm{VS}}\rangle\approx200\,\upmu\mathrm{eV}$ (for large negative $F$) to $\langle E_{\mathrm{VS}}\rangle > 1\,\mathrm{meV}$ (for large positive $F$).
The point of maximum sensitivity is near the design field strength $F_{\mathrm{opt}}$, see Fig.~\ref{fig:mod_WW-electric field dep}\,(b).
This pronounced field dependency is explained by the different confinement behavior of the electronic wave function shown in Fig.~\ref{fig:mod_WW-electric field dep}\,(c).
At large negative $F$, the envelope wave function does not probe
the Ge-rich modulations, such that the structure effectively behaves
like a conventional Si/SiGe QW with smooth interfaces.
In the opposite case, at large positive field strength, the envelope overlaps dominantly with the most Ge-rich segments only, leading to a large valley splitting expectation value.
Over the entire range of electric fields, the deterministic
component dominates over the disorder-induced contribution to the
valley splitting.
We believe that this enhanced tunability could be useful to switch from high to low valley splitting regimes on demand,
\emph{e.g.}, in order to control the rate of phonon-assisted valley
relaxation \cite{Langrock2023, Losert2024, Oda2026} to ensure a definite valley state prior to the execution of two-qubit gates.
Moreover, the modulated wiggle well design could be relevant for the tuning of energy splittings of hybrid qubits \cite{Shi2012, Kim2015}.
To achieve wide tunability of the vertical electric field, advanced device designs with back-gate electrodes are advantageous \cite{Marcogliese2026}.

\subsection{Narrow Well}

When optimizing for the reduction of the disorder-induced component (C), we obtain the epitaxial profile shown in Fig.~\ref{fig: optim results}\,(c).
The second column shows the envelope wave function of the ground state electron, which makes clear how the reduction of disorder is achieved:
By accumulating the Ge atoms in a plateau near the bottom QW interface, the structure becomes effectively a narrow QW, where the electron is localized in a narrow Ge-free segment.
In combination with additional smoothing of the upper interface, this leads to a reduction of the overlap with Ge atoms and therefore to a reduction of valley splitting
volatility (see also the complex plane distribution in the fourth
column).
The narrow well provides only a moderate deterministic
valley splitting, such that the deterministic component ratio is small compared to the modulated wiggle well ($Q=0.85$, $\zeta=0.9$).

\begin{figure*}
\includegraphics[width=0.94\textwidth]{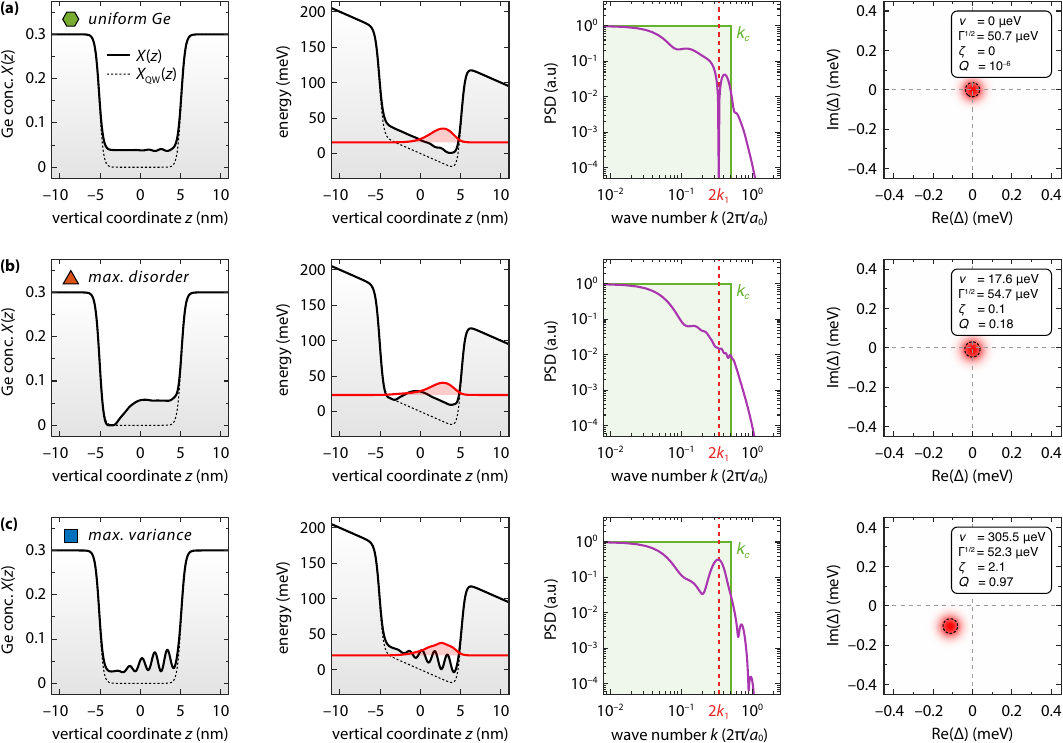}
\caption{Optimization under inverted objectives. \textbf{(a)}~Minimization of the deterministic component (D) leads to a (nearly) uniform Ge distribution in the QW.
The PSD (third column) shows a sharp dropout around $2k_{1}$, such that the resulting valley splitting is fully disorder-induced.
\textbf{(b)}~Maximization of the disorder-induced component (E) of the intervalley coupling matrix element yields a profile similar to the uniform Ge distribution, but with an increased Ge content in the region that overlaps with the envelope wave function.
\textbf{(c)}~Maximization of the valley splitting variance (F) leads to a wiggle-well type structure with non-zero baseline Ge concentration.
\label{fig: optim inverted}
}
\end{figure*}

\subsection{Germanium Spike}

We return to the maximization of the deterministic component (A),
but with a reduced cutoff wave number $k_{c}=0.07\times2\pi/a_{0}$
in order to suppress the direct $2k_{1}$ resonance.
In this case, our optimization procedure yields a single Ge peak at about $3\,\mathrm{nm}$ below the upper interface, see Fig.~\ref{fig: optim results}\,(d).
The structure allows for a significant deterministic enhancement due to a higher-order effect, where the combination of the heterostructure potential and the envelope wave function boosts the $2k_{1}$-resonance (see the PSD in the third column).
On the other hand, the strong overlap with Ge-rich domains leads to a large disorder-induced component, too.
More details on the underlying mechanism are given in Ref.~\cite{Thayil2025}.
We remark that the structure has some similarity with the ``Ge spike'' described in Refs.~\cite{McJunkin2021, Losert2023, Salamone2026}, which is why we adopt this notion here.
Finally, the valley splitting in the Ge spike has a similar field tunability as the modulated wiggle well, but with a smaller range of achievable mean valley splittings, see Fig.\,\ref{fig:mod_WW-electric field dep}\,(a)--(b).
The underlying mechanism is shown in Fig.\,\ref{fig:mod_WW-electric field dep}\,(d), which resembles a tunable double dot system with strongly different confinement potentials near the top and bottom interface.

\subsection{Optimization of Inverted Objectives}
It is illustrative to carry out the epitaxial profile optimization
also under inversion of the objective functions to investigate solutions
that enhance undesirable properties.
Specifically, we consider optimization
objectives that are inverse to  $J_{0}^{\left(A\right)}\left(x,\psi_{0}\right)$
and $J_{0}^{\left(C\right)}\left(x,\psi_{0}\right)$ to obtain structures
with either very low deterministic or highly volatile valley splittings. In addition, we investigate the maximization of the variance of the valley splitting (spread of the Rice distribution), see Eq.~\eqref{eq: variance E_VS}.
We consider:
\begin{enumerate}[label=(\Alph*), start=4]
\item Minimization of deterministic component
\begin{equation*}
J_{0}^{\left(D\right)}\left(x,\psi_{0}\right)=\frac{1}{J_{0}^{\left(A\right)}\left(x,\psi_{0}\right)}=\frac{\nu\left(x,\psi_{0}\right)}{E_{\mathrm{ref}}}.
\end{equation*}
\item Maximization of disorder-induced random component
\begin{equation*}
J_{0}^{\left(E\right)}\left(x,\psi_{0}\right)=\frac{1}{J_{0}^{\left(C\right)}\left(x,\psi_{0}\right)}=\frac{E_{\mathrm{ref}}}{\sqrt{2\Gamma\left(x,\psi_{0}\right)}}.
\end{equation*}
\item Maximization of the valley splitting variance
\begin{equation*}
J_{0}^{\left(F\right)}\left(x,\psi_{0}\right)=\frac{E_{\mathrm{ref}}}{\sqrt{\mathrm{Var}\left(E_\mathrm{VS}\left(x,\psi_{0}\right)\right)}}.
\end{equation*}
\end{enumerate}
The numerical results for a limited Ge budget $x_{\mathrm{Ge}}=5\%$
and cutoff wave number $k_{c}=0.5\times2\pi/a_{0}$ are shown in Fig.~\ref{fig: optim inverted}.

For the minimization of the deterministic component (D) we obtain
a (nearly) uniform Ge concentration in the QW, see Fig.~\ref{fig: optim inverted}\,(a).
The combination of the confinement potential and the envelope wave
function leads to a PSD that features a sharp dropout around wave
number $2k_{1}$, such that contributions to the deterministic
enhancement of the valley splitting are strongly suppressed.
The valley splitting is thus \emph{fully} disorder-induced and has no deterministic enhancement at all.
While the uniform Ge concentration leads to an enhancement
of the mean valley splitting in comparison to the pure silicon QW
(with the same interface width), this enhancement comes along with
a large volatility.
As a consequence, numerous spin-valley hotspots are to be expected in these structures.
Quantum wells with uniform Ge concentration
profiles have been studied in Refs.~\cite{PaqueletWuetz2022, Losert2023, Losert2024}.

The maximization of the disorder-induced random component (E)
leads to an epitaxial structure similar to the uniform Ge distribution, but with an enhanced Ge accumulation in the region that overlaps with the envelope wave function, see Fig.~\ref{fig: optim inverted}\,(b).
In order to respect the limited Ge budget constraint, the Ge concentration is reduced near the lower interface to provide additional Ge atoms in the overlap region with the electronic envelope wave function.
The epitaxial structure is similar to a mirror inverse of the narrow well profile shown in Fig.~\ref{fig: optim results}\,(c),
which is interesting as the optimization objective $J_{0}^{\left(E\right)}$ is the inverse of $J_{0}^{\left(C\right)}$.
The magnitude of the disorder-induced component is, however, only slightly larger than in case (D) or the Ge spike in Fig.~\ref{fig: optim results}\,(d).

Finally, we consider maximization of the valley splitting variance (F), which differs from maximization of the disorder-induced component as the spread of the Rice distribution can become larger at higher mean values. The result is shown in Fig.~\ref{fig: optim inverted}\,(c), which is a wiggle-well type profile but with non-zero baseline Ge concentration. As in cases (A) and (B), the wiggle well structure induces a vertically stretched envelope wave function that is extended over almost the entire QW domain.
This way, a large overlap with disorder-inducing Ge atoms is achieved, leading to a large variance of the valley splitting.
The magnitudes of the deterministic and disorder-induced components are very similar to those of the Ge spike, see Fig.~\ref{fig: comparison map}, which underpins again the large uncertainty of the valley splitting in the Ge spike heterostructure.

\section{Summary and Outlook} \label{sec: summary}

\begin{figure}
\includegraphics[width=1\columnwidth]{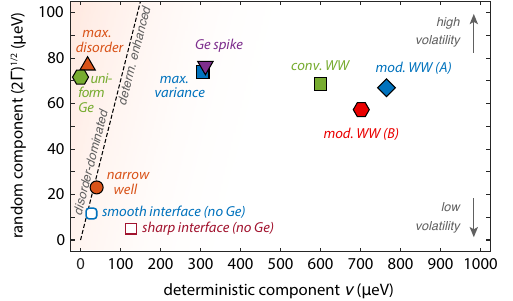}

\caption{Comparison of deterministic and disorder-induced components for the
different heterostructures obtained by variational optimization with
limited Ge budget $x_{\mathrm{Ge}}=5\%$. The dashed line indicates
$\nu/\sqrt{2\Gamma}=0.3507$, which separates the disorder-dominated
from the deterministically enhanced regime ($Q=0.5$). For comparison,
the plot shows also the values of a smooth ($\sigma_{u,l}=0.5\,\mathrm{nm}$)
and a sharp ($\sigma_{u,l}=0\,\mathrm{nm}$) interface with no Ge
included in the QW.}

\label{fig: comparison map}
\end{figure}

We have developed a variational optimization approach for the computation of Ge concentration profiles in Si/SiGe QWs to enhance the valley splitting in gate-defined QDs.
We investigated a number of different optimization objectives, including maximization of the deterministic
component, minimization of the disorder-induced component and optimal balancing of both contributions (reliable enhancement).
Our method has recovered heuristic heterostructure designs previously described in the literature as special solutions of a constrained optimization problem (\emph{e.g.}, wiggle wells, Ge spike, narrow well, uniform Ge concentration). 
The characteristics of the various profiles are summarized in Fig.~\ref{fig: comparison map}, which shows the typical
deterministic and disorder-induced contributions to the valley splitting expected for the structures with a total Ge content of 5\% in the
QW domain.
For comparison, the plot also shows the results for conventional $\mathrm{Si}_{0.7}\mathrm{Ge}_{0.3}/\mathrm{Si}/\mathrm{Si}_{0.7}\mathrm{Ge}_{0.3}$ QWs (with no Ge in the QW domain) with smooth and sharp interfaces.
The results of the optimization objectives (A) and (B) are far in the deterministically enhanced regime, while the result of (C) offers only a weak deterministic enhancement. 
The optimization for the inverted objectives (D) and (E) leads to structures deep inside the disorder-dominated regime.

Our main result is the \emph{modulated wiggle well}, a modification of the conventional sinusoidal wiggle-well profile that is adapted to the electric-field-induced asymmetry of the electronic ground-state confinement in the QW.
We have shown that the modulated wiggle well outperforms the conventional wiggle well by offering both a larger deterministic component and a reduced disorder-related contribution to the intervalley coupling matrix element (\emph{i.e.}, it provides a more reliable enhancement), see Fig.\,\ref{fig: comparison map}.
These properties are crucial to ensure a high degree of reproducibility of qubits across devices required for a scalable quantum computing technology platform.
Moreover, the modulated wiggle well shows a strong dependency on the applied electric field, which allows for a wide tunability range of the mean valley splitting
from $200\,\upmu\mathrm{eV}$ to more than $1\,\mathrm{meV}$.
We believe that these enhanced prospects for tunability can be useful for the design of \emph{switchable} spin qubits with on-demand adjustable valley splitting, \emph{e.g.}, to control the rate of relaxation processes prior to quantum gate operations or readout. 

Finally, we emphasize that, since our work is based on a simplified model of the SiGe heterostructure, it leaves out several aspects that are important for spin qubits.
These include, \emph{e.g.}, the influence of the epitaxial profile and alloy disorder on the spin-orbit coupling strength, the effective spin-valley coupling, and $g$-factor modifications. 
Likewise, the effects of bond-length and bond-angle variations in the disordered heterostructure were not taken into account.
Future work on epitaxial profile optimization should therefore incorporate more comprehensive theoretical models to account for these additional spin and disorder-related effects.

\section*{Code Availability}
MATLAB code to reproduce the simulation results and
figures of this work is available on GitHub \cite{Thayil2025c}.

\begin{acknowledgments}
This work was funded by the Deutsche Forschungsgemeinschaft (DFG,
German Research Foundation) under Germany's Excellence Strategy -{}-
The Berlin Mathematics Research Center MATH+ (EXC-2046/1, project
ID: 390685689) and the Cluster of Excellence ``Matter and Light for
Quantum Computing'' (ML4Q, EXC 2004/2, project ID 390534769). M.K.
acknowledges helpful discussions with Shalva Amiranashvili.
\end{acknowledgments}

\appendix

\section{Mean, Covariance and Pseudo-Covariance \label{sec: formulas for Delta}}

Following \cite{Thayil2025}, we employ a factorization ansatz for
the ground state envelope wave function
\begin{equation*}
\Psi_{0}\left(\mathbf{r}\right) =\phi_{0}\left(x,y\right)\psi_{0}\left(z\right)
\end{equation*}
separating the problem in vertical and in-plane directions. The expressions
for the deterministic valley splitting (\ref{eq: Delta det-1}) and
the covariance and pseudo-covariance (\ref{eq: Gamma and C}) are
evaluated using the ground state wave function of the two-dimensional QD
\begin{equation*}
\phi_{0}\left(x,y\right) =\frac{1}{\pi^{1/4}l_{x}^{1/2}}\mathrm{e}^{-\frac{1}{2}\left(x/l_{x}\right)^{2}}\times\frac{1}{\pi^{1/4}l_{y}^{1/2}}\mathrm{e}^{-\frac{1}{2}\left(y/l_{y}\right)^{2}}
\end{equation*}
where $l_{i}^{2}=\hbar/\left(m_{t}\omega_{i}\right)$, $i\in\left\{ x,y\right\} $.
With this, we derive expressions for the deterministic
component
\begin{align}
\Delta_{\mathrm{det}}\left(x,\psi_{0}\right) & =\sum_{n}C_{n}^{\left(2\right)}\mathrm{e}^{-\left(\frac{nG_{0,x}l_{x}}{2}\right)^{2}}\mathrm{e}^{-\left(\frac{nG_{0,y}l_{y}}{2}\right)^{2}}\times\label{eq: Delta det (only z)}\\
 & \phantom{=}\int\mathrm{d}z\,\mathrm{e}^{-i\left(nG_{0,z}+2k_{0}\right)z}\bigg(U_{\mathrm{QW}}\left(z\right)+U_{F}\left(z\right)+\nonumber \\
 & \hphantom{=\int\mathrm{d}z\,}+\frac{\hbar\omega_{x}}{2}\left[\frac{1}{2}-\left(\frac{nG_{0,x}l_{x}}{2}\right)^{2}\right]\nonumber \\
 & \hphantom{=\int\mathrm{d}z\,}+\frac{\hbar\omega_{y}}{2}\left[\frac{1}{2}-\left(\frac{nG_{0,y}l_{y}}{2}\right)^{2}\right]\bigg)\psi_{0}^{2}\left(z\right)\nonumber 
\end{align}
and the covariance
\begin{align*}
\Gamma\left(x,\psi_{0}\right) & =\frac{\left(\Delta E_{c}\right)^{2}\Omega_{a}}{2\pi l_{x}l_{y}}\sum_{n}C_{n}^{\left(4\right)}\mathrm{e}^{-\left(\frac{nG_{0,x}l_{x}}{2\sqrt{2}}\right)^{2}}\mathrm{e}^{-\left(\frac{nG_{0,y}l_{y}}{2\sqrt{2}}\right)^{2}} \\
 & \phantom{=}\times\int\mathrm{d}z\,\mathrm{e}^{-inG_{0,z}z}X\left(z\right)\left(1-X\left(z\right)\right)\psi_{0}^{4}\left(z\right).
\end{align*}
The pseudo-covariance is obtained as
\begin{align*}
C\left(x,\psi_{0}\right) & =\frac{\left(\Delta E_{c}\right)^{2}\Omega_{a}}{2\pi l_{x}l_{y}}\sum_{n}D_{n}^{\left(4\right)}\mathrm{e}^{-\left(\frac{nG_{0,y}l_{y}}{2\sqrt{2}}\right)^{2}}\mathrm{e}^{-\left(\frac{nG_{0,x}l_{x}}{2\sqrt{2}}\right)^{2}}\\
 & \phantom{=}\times\int\mathrm{d}z\,\mathrm{e}^{-i\left(nG_{0,z}+4k_{0}\right)z}X\left(z\right)\left(1-X\left(z\right)\right)\psi_{0}^{4}\left(z\right).
\end{align*}
The Bloch factor coefficients in the above equations are
\begin{equation*}
C_{n}^{\left(2\right)}  =\sum_{\mathbf{G},\mathbf{G}'}c_{+}^{*}\left(\mathbf{G}\right)c_{-}\left(\mathbf{G}'\right)\delta_{\mathbf{G}-\mathbf{G}',n\mathbf{G}_{0}}
\end{equation*}
with
\begin{equation*}
\mathbf{G}_{0} =\left(I-\varepsilon\right)\left(\mathbf{b}_{1}+\mathbf{b}_{2}\right)=\frac{4\pi}{a_{0}}\left(\begin{array}{c}
-\varepsilon_{z,x}\\
-\varepsilon_{y,z}\\
1-\varepsilon_{z,z}
\end{array}\right).
\end{equation*}
In this paper, we use $\varepsilon_{z,x} = \varepsilon_{y,z}=0$ and $\varepsilon_{z,z} =\varepsilon_{\perp}$, see Sec.~\ref{sec:Statistical-Distribution}.
The coefficients for the covariance and pseudo-covariance are obtained
from the set of $C_{n}^{\left(2\right)}$ coefficients as
\begin{align*}
C_{n}^{\left(4\right)} & =\sum_{m}C_{m-n}^{\left(2\right)*}C_{m}^{\left(2\right)},\\
D_{n}^{\left(4\right)} & =\sum_{m}C_{m}^{\left(2\right)}C_{n-m}^{\left(2\right)}.
\end{align*}
Due to the rapidly oscillating factor $\exp\left(-4ik_{0}z\right)$
in the integrand, the pseudo-covariance is typically much smaller
than the covariance $\left|C\right|\ll\Gamma$. Hence, the pseudo-covariance
is ignored in most of the computations by employing the circular approximation,
see Sec.~\ref{sec:Statistical-Distribution}.

\section{Conventional Wiggle Well}
The free-shape optimization results
are compared with the conventional sinusoidal (long-period)
wiggle well given as
\begin{equation*}
x_{\mathrm{ww}}\left(z\right) =A\,\sin^{2}\left(k_{1}z\right)\,\Xi\left(z\right),
\end{equation*}
where $A$ is the Ge concentration amplitude.
The indicator function $\Xi\left(z\right)$
restricts the wiggle well to the QW domain. For a fair
comparison, we choose the amplitude $A$ such that the overall
Ge concentration matches the target Ge budget. Hence, the amplitude is chosen such that
\begin{equation*}
\overline{X}\left(x_{\mathrm{ww}}\left(A\right)\right)-x_{\mathrm{Ge}}=0.
\end{equation*}
The function to quantify the mean Ge content $\overline{X}\left(x\right)$ of a particular profile
is given in Eq.~(\ref{eq: mean Ge content}). The equation is
solved for $A$ numerically.
For $x_{\mathrm{Ge}}=5\%$ and an $h=75\,\mathrm{ML}$ QW, we obtain $A=0.0764$.


%

\end{document}